\documentclass[british]{article}
\usepackage[T1]{fontenc}
\usepackage{textcomp}
\usepackage[utf8]{inputenc}
\usepackage{babel}
\usepackage{url}
\usepackage{amsmath}
\usepackage{amssymb}
\usepackage{graphicx}
\usepackage[a4paper]{geometry}
\usepackage[pdfusetitle,
 bookmarks=true,bookmarksnumbered=false,bookmarksopen=false,
 breaklinks=false,pdfborder={0 0 1},backref=false,colorlinks=false]
 {hyperref}

\makeatletter
\usepackage{amsmath}
\usepackage{graphicx}
\usepackage[figure]{hypcap}
\hypersetup{hidelinks}

\makeatother

\begin{document}
\title{Depth-resolved elastic property tomography using higher-order shear-horizontal
modes}
\author{Peter Huthwaite}
\maketitle
\begin{center}
Department of Mechanical Engineering, Imperial College London, London,
United Kingdom
\par\end{center}

\section*{Abstract}

Early material degradation can produce weak, local changes in elastic
stiffness before a crack or void forms, but conventional ultrasonic
measurements dilute these changes through path and thickness averaging.
This paper introduces DEPTH (Depth-resolved Elastic Property Tomography
using Higher-order modes), which uses mode-to-mode scattering among
higher-order guided waves to reconstruct elastic-property variations
as a function of both axial position and depth through a plate. The
formulation is developed and evaluated here using shear-horizontal
(SH) modes, for which the reconstructed property is shear modulus.
A Born formulation shows that each incident-received mode pair contains
at most two cosine depth orders, allowing the through-thickness field
to be recovered from sufficiently diverse modal mixtures across mode
pairs and frequency. The resulting matrix-free regularised inverse
is evaluated progressively using direct two-dimensional modal data,
practical single-surface excitation and reception, and full three-dimensional
finite-element data. In a 10 mm plate, the ideal two-dimensional case
gives a through-thickness point-spread FWHM of 0.84-1.03 mm, while
the single-surface configuration recovers four compact targets at
the correct depths from transmission between two 20-element arrays.
This provides the first demonstration of depth-resolved elastic-property
tomography from multimode guided-wave transmission with excitation
and reception confined to a single physical plate surface. With full
three-dimensional propagation and finite-width transduction, the reconstruction
retains the compact-target positions with a whole-image correlation
of 0.899 against a matched acquisition-limited reconstruction. For
the cross-mode surface datasets, the recovered contrast is relative
to its local through-thickness mean. Independent additive measurement
noise produces gradual degradation, with a median maximum localisation
error of 0.150 mm at 30 dB SNR; an isolated assumed-thickness perturbation
of magnitude up to 2\% after modal-dispersion calibration leaves the
recovered depth-varying coefficients largely unchanged, with whole-image
correlation above 0.971. These simulation results demonstrate that
higher-order guided-wave mode conversion can provide local depth-resolved
elastic-property tomography from surface-access measurements, which
could be critical for characterising early-stage damage such as fatigue.

\section{Introduction}

Many forms of material degradation begin before there is a crack or
void to detect. At these early stages, the changes can be subtle,
often manifesting as changes in the elastic stiffness of the material.
This has been observed for a range of degradation mechanisms, including
fatigue \cite{sarrisUltrasonicMethodsDetection2023}, creep \cite{gomezalvarez-arenasUltrasonicEvaluationCreep1993},
hydrogen embrittlement \cite{seniorUltrasonicDetectionHydrogen1984},
thermal ageing \cite{dearaujofreitasNondestructiveCharacterizationEvaluation2011}
and irradiation embrittlement \cite{ishiiDevelopmentNondestructiveTesting2002}.
In particular, early-stage fatigue has been shown to change ultrasonic
wave speed, linked to changes in elastic modulus and the underlying
change in dislocation state \cite{sarrisFatigueStateCharacterization2023}.

Ultrasound is attractive for this problem because wave propagation
depends directly on the elastic properties of the material, allowing
changes in material state to be observed through changes in wave speed,
attenuation and scattering. These measurements can probe through the
volume of a component and do not require a crack or void to have formed
before a change can be observed. This provides a route to detecting
degradation earlier, while it is still expressed as a material-property
change rather than as a discrete defect. This has direct value for
asset management: detecting degradation at this stage increases the
margin available for maintenance decisions, allowing work to be aligned
with scheduled outages where appropriate while avoiding both unsafe
continued operation and unnecessarily early intervention.

These material properties are typically a challenge to identify: they
are subtle, but also localised such that the majority of ultrasound
propagation typically occurs outside the damage region, giving poor
sensitivity and specificity because most of the measured response
comes from undamaged material along the propagation path \cite{sarrisUltrasonicMethodsDetection2023}.
Sarris et al.~demonstrated this for near-surface fatigue damage:
confining the measurement closer to the affected region produced a
substantially larger velocity change than a through-thickness measurement
\cite{sarrisUltrasonicMethodsDetection2023}. This illustrates the
sensitivity penalty associated with averaging a local material-property
change through the full wall thickness. The problem is therefore not
simply measurement precision: the measurement itself is weakly sensitive
to the region of interest.

The central measurement problem is therefore to localise the measurement
sensitivity through the component thickness, rather than simply to
measure the same through-thickness average more precisely. Without
this localisation, a small region of degraded material makes only
a small contribution to the measured response, however accurately
that response is measured. Depth resolution changes the measurement
itself: it enables the local property change to be separated from
the undamaged material that otherwise dominates the response.

Existing methods have addressed depth sensitivity, multimode characterisation
and spatial imaging in different forms. Depth-dependent elastic-property
reconstruction has been demonstrated in related inverse problems.
In geophysics, Rayleigh-wave dispersion has been inverted to obtain
near-surface shear-wave velocity profiles as a function of depth \cite{xiaEstimationNearsurfaceShearwave1999},
while ultrasonic surface-wave and elastic-wave approaches have reconstructed
continuously varying through-thickness properties in functionally
graded materials \cite{luMechanicalPropertiesEstimation2011,liuMaterialCharacterizationFGM2001}.
These approaches recover a through-thickness profile by exploiting
depth-dependent wave sensitivity, but generally assume that the material
properties vary only with depth over the measurement region. Lateral
localisation is therefore absent by construction. Multimodal Lamb-wave
dispersion inversion similarly combines information from several modes
to estimate plate thickness and bulk elastic constants, but remains
a laterally averaged parameter-estimation problem \cite{chenHighresolutionLambWaves2021}.
Within guided-wave NDE, spatial variations in stiffness have been
imaged using diffraction tomography based on Mindlin plate theory
and guided-wave inversion \cite{roseMindlinPlateTheory2010a,ratasseppQuantitativeImagingYoungs2018a},
but these methods are depth-insensitive: they reconstruct plate-theory
stiffness parameters or an average through-thickness Young's modulus
rather than the depth of a local stiffness change. This is the complementary
limitation: the reconstructed property can vary along the component,
but not independently through its thickness. Multimode Lamb-wave tomography
and guided-wave full-waveform inversion have improved in-plane reconstruction
by exploiting complementary modal dispersion or more complete waveform
information \cite{huangMultiModeElectromagneticUltrasonic2016a,raoGuidedWaveTomography2016a}.
In these plate-imaging formulations, however, the reconstructed field
is an in-plane thickness or effective wave-speed map rather than a
material property that varies independently through the thickness.
Higher-order SH modes have also been used to quantify adhesive thickness,
interfacial shear stiffness and adhesive shear modulus in a prescribed
layered joint \cite{koodalilQuantifyingAdhesiveThickness2021}, and
guided SH-wave scattering has been used to reconstruct the upper and
lower boundaries of an internal cavity \cite{wangInverseShapeReconstruction2015}.
These studies exploit depth-dependent modal information, but estimate
a small number of prescribed layer, interface or boundary parameters
rather than isolating a weak material-property change at an unknown
depth. Mode-converted Lamb and SH responses are also known to vary
strongly with part-thickness defect depth, but such approaches have
predominantly targeted detection, localisation or prescribed geometry
rather than a distributed depth-resolved property field \cite{liModeConversionFundamental2024}.
Forward studies further show that SH waves are sensitive to arbitrary
continuous through-thickness material inhomogeneity \cite{shuvalovShearHorizontalWaves2008}.
Born scattering in waveguides has also been related explicitly to
constrained spatial-frequency sampling, with the free-space Ewald
sphere replaced by a waveguide Ewald strip \cite{ghoshroyBornScattererAcoustical2003}.
More directly, Bonnetier et al.~reconstructed a distributed scalar
inhomogeneity varying in both waveguide coordinates from multifrequency
reflection data using a Born/Fourier inverse \cite{bonnetierSmallDefectsReconstruction2022}.
In that formulation, however, a prescribed incident guided mode is
assumed and the measured field is available across the complete waveguide
cross-section; experimental multimode Lamb-wave tomography has instead
used mode-selective A0 and S0 EMAT arrays \cite{huangMultiModeElectromagneticUltrasonic2016a}.
Neither approach recovers a multimode incident-to-received transfer
matrix from mixed point-source and point-receiver measurements confined
to a single accessible plate surface. The present study instead reconstructs
a multimode incident-to-received transmission matrix from transmitter
and receiver apertures that are separated axially but confined to
the same physical plate surface. The resulting single-surface multimode
transmission configuration therefore provides the first demonstration
of depth-resolved elastic-property tomography while requiring access
to only one physical plate surface.

This paper introduces DEPTH - Depth-resolved Elastic Property Tomography
using Higher-order modes - which uses a family of higher-order guided-wave
modes to reconstruct a tomographic image of elastic-property variations
as a function of both position along a component and depth through
its thickness. This changes the nature of the measurement: rather
than reducing a local property change to a path- or thickness-averaged
response, DEPTH reconstructs where within the thickness the change
occurs and its magnitude. Here, depth-resolved means that through-thickness
position is an independent reconstruction coordinate, rather than
being represented only by a local thickness, layer or other prescribed
depth parameter. The purpose of the depth resolution is therefore
not simply to locate the change; it directly increases sensitivity
to the local material state by separating it from the surrounding
undamaged material. In the SH formulation developed here, multiple
incident and received modes provide mixtures of difference and sum
depth orders at axial spatial frequencies set by modal wavenumber
differences. The single-surface implementation recovers the required
multimode transmission data through synthesis and calibration from
point-source and point-receiver measurements on that same physical
plate surface.

The formulation is developed in Section~\ref{sec:Methodology}, which
makes explicit how each mode pair encodes a small set of through-thickness
orders and how these are combined in the regularised inverse. Section~\ref{sec:2d-evaluation}
then carries the same measurement representation into progressively
more practical two-dimensional tests: direct finite-element modal
data establish the attainable depth resolution, before a practical
single-surface excitation and reception model tests whether the required
modal information can be recovered from an accessible surface. Section~\ref{sec:3d-evaluation}
extends the evaluation to full three-dimensional simulation, including
finite-width transduction and transverse target extent. Section~\ref{sec:sensitivity}
examines sensitivity to measurement noise and uncertainty in assumed
wall thickness, before conclusions are drawn in Section~\ref{sec:conclusions}.

\section{Methodology\label{sec:Methodology}}

\subsection{Reference plate and SH modes}

DEPTH reconstructs a spatial perturbation in shear modulus from calibrated
mode-to-mode SH scattering measurements. The reference plate occupies
$0<z<h$ and is isotropic, with shear modulus $\mu_{0}$ and density
$\rho_{0}$. The formulation assumes that the density is unchanged
and defines the fractional shear-modulus perturbation as
\begin{equation}
q(x,z)=\frac{\mu(x,z)-\mu_{0}}{\mu_{0}}.\label{eq:q}
\end{equation}
A time dependence $\exp(-\mathrm{i}\omega t)$ is used. The reference
waveguide is assumed flat and nominally homogeneous and isotropic;
the present operator does not include material attenuation, curvature
or background anisotropy. In the reference plate the SH equation is
\begin{equation}
\nabla\cdot(\mu_{0}\nabla u)+\rho_{0}\omega^{2}u=0.\label{eq:shWaveEq}
\end{equation}
The traction-free surfaces require $\partial u/\partial z=0$ at $z=0,h$.
The normalised SH mode shapes and axial wavenumbers are therefore
\begin{equation}
\phi_{m}(z)=N_{m}\cos(\alpha_{m}z),\qquad\alpha_{m}=\frac{m\pi}{h},\qquad k_{m}=\left(\frac{\rho_{0}\omega^{2}}{\mu_{0}}-\alpha_{m}^{2}\right)^{1/2},\label{eq:modes}
\end{equation}
where $N_{0}=h^{-1/2}$ and $N_{m}=(2/h)^{1/2}$ for $m>0$. Only
modes with real positive $k_{m}$ are treated as propagating measurement
channels.

\subsection{Linearised mode-pair measurement}

Write the total field as the sum of incident and scattered fields,
$u=u^{\mathrm{i}}+u^{\mathrm{s}}$. Define the reference operator
for the homogeneous reference plate as
\begin{equation}
L_{0}u=\nabla\cdot(\mu_{0}\nabla u)+\rho_{0}\omega^{2}u.\label{eq:referenceoperator}
\end{equation}
The incident mode $u_{m}^{\mathrm{i}}$ is a solution of the reference
problem, so $L_{0}u_{m}^{\mathrm{i}}=0$. Substitution of $\mu=\mu_{0}(1+q)$
into the wave equation \eqref{eq:shWaveEq}, while using the fact
that the incident field satisfies the reference equation, gives
\begin{equation}
L_{0}u^{\mathrm{s}}+\nabla\cdot\left(\mu_{0}q\nabla u_{m}^{\mathrm{i}}\right)+\nabla\cdot\left(\mu_{0}q\nabla u^{\mathrm{s}}\right)=0.\label{eq:bornexpanded}
\end{equation}
The final term contains the perturbation acting on the already scattered
field and is second order in the scattering process. The first Born
approximation neglects this term, leaving
\begin{equation}
L_{0}u^{\mathrm{s}}=-\nabla\cdot\left(\mu_{0}q\nabla u_{m}^{\mathrm{i}}\right)\equiv f_{\mathrm{B}}.\label{eq:born}
\end{equation}
The right-hand side of Equation \eqref{eq:born}, denoted $f_{\mathrm{B}}$,
acts as a distributed secondary source generated by the interaction
of the incident strain $\nabla u_{m}^{\mathrm{i}}$ with the modulus
perturbation $\mu_{0}q$. For received mode $n$, introduce the reciprocal
field $v_{n}=B_{n}\phi_{n}(z)e^{-\mathrm{i}k_{n}x}$, obtained by
launching the received mode reciprocally from the receiver towards
the scattering region. The reciprocity relation then expresses the
complex amplitude scattered from incident mode $m$ into received
mode $n$ as the overlap of this reciprocal field with $f_{\mathrm{B}}$.
Integrating by parts transfers the divergence onto the known reciprocal
field and gives
\begin{equation}
\widetilde{d}_{nm}=\widetilde{\kappa}_{nm}\int\!\!\int q(x,z)S_{nm}(z)e^{\mathrm{i}(k_{m}-k_{n})x}\,\mathrm{d}z\,\mathrm{d}x.\label{eq:rawmeasurement}
\end{equation}
Here, $\widetilde{d}_{nm}$ is the complex scattering amplitude for
incident mode $m$ and received mode $n$. The source, receiver, modal
normalisation and calibration factors that are not part of the spatial
sensitivity are collected into the complex coefficient $\widetilde{\kappa}_{nm}$.
The through-thickness sensitivity function is
\begin{equation}
S_{nm}(z)=k_{m}k_{n}\phi_{m}(z)\phi_{n}(z)+\phi_{m}'(z)\phi_{n}'(z).\label{eq:kernel}
\end{equation}
The first term comes from the product of the axial derivatives of
the incident and reciprocal fields and therefore represents axial
shear strain; the second comes from their through-thickness derivatives.
Appendix \ref{sec:appendix-kernel} gives the full derivation from
the perturbed wave equation \eqref{eq:born} to \eqref{eq:rawmeasurement}
and \eqref{eq:kernel}.

The reconstruction uses the axial spatial-frequency convention
\begin{equation}
K_{x}=k_{n}-k_{m},\label{eq:kx}
\end{equation}
with a positive Fourier exponent. Equation \eqref{eq:rawmeasurement}
contains $e^{-\mathrm{i}K_{x}x}$. Since $q$ is real, the complete
calibrated equation can be conjugated once when the measurement set
is constructed, so that $d_{nm}=\widetilde{d}_{nm}^{*}$ and $\kappa_{nm}=\widetilde{\kappa}_{nm}^{*}$.
This sign conversion is applied to both the data and the complex calibration
coefficient, rather than being handled inside the inverse. The measurement
equation used from this point onwards is therefore
\begin{equation}
d_{nm}=\kappa_{nm}\int\!\!\int q(x,z)S_{nm}(z)e^{\mathrm{i}K_{x}x}\,\mathrm{d}z\,\mathrm{d}x.\label{eq:canonicalmeasurement}
\end{equation}

\subsection{Through-thickness encoding by mode pairs}

Substituting the cosine mode shapes into \eqref{eq:kernel} and using
the product-to-sum identities gives
\begin{equation}
S_{nm}(z)=\frac{N_{m}N_{n}}{2}\left[(k_{m}k_{n}+\alpha_{m}\alpha_{n})\cos\left(\frac{(m-n)\pi z}{h}\right)+(k_{m}k_{n}-\alpha_{m}\alpha_{n})\cos\left(\frac{(m+n)\pi z}{h}\right)\right].\label{eq:twoorders}
\end{equation}
Because cosine is even, a single incident-received pair is sensitive
to no more than two cosine depth orders,
\[
p=|m-n|,\qquad p=m+n.
\]
This is the source of the depth information in DEPTH. Different mode
pairs do not measure depth directly; they provide different mixtures
of a sparse set of through-thickness spatial orders. Under the Fourier
diffraction theorem for a Born scattering problem in unbounded space,
a source-receiver combination samples a particular spatial-frequency
component of the object. Confinement by the plate quantises the through-thickness
dependence into discrete modal orders, so a mode-pair measurement
instead superposes at most two depth orders at the same axial K-space
coordinate. The inverse must therefore separate these known mixtures
using measurements from multiple mode pairs and frequencies. 

\begin{figure}
\includegraphics[width=1\textwidth]{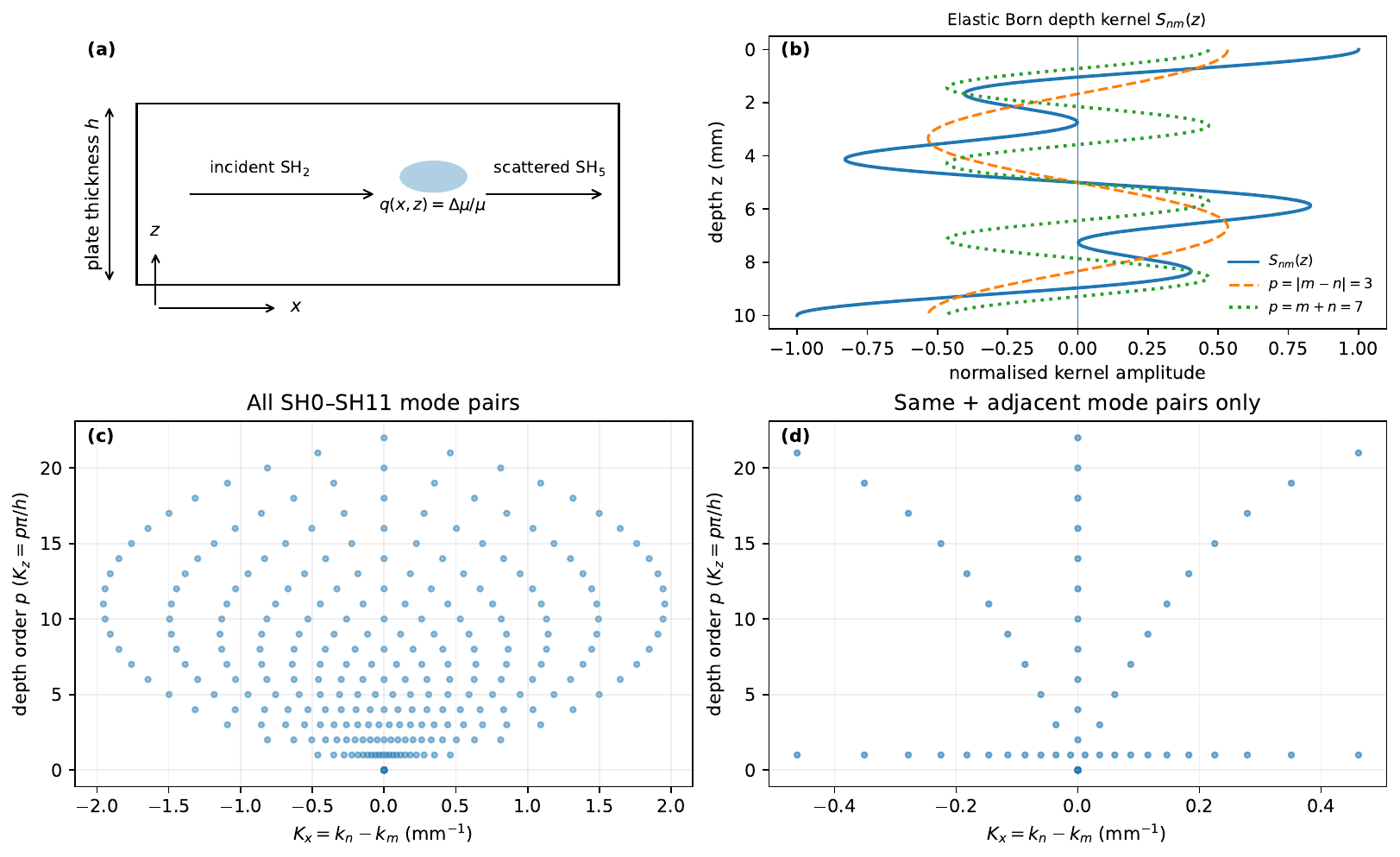}\caption{DEPTH principle and modal sampling. (a) An incident SH2 mode scatters
from a local shear-modulus perturbation into SH5. (b) The corresponding
elastic Born depth kernel contains only the two depth orders p=|m-n|=3
and p=m+n=7. (c) Axial spatial frequencies and depth orders sampled
by all SH0-SH11 mode pairs at 2.0 MHz. (d) Equivalent sampling at
2.0 MHz using same- and adjacent-mode pairs only.\label{fig:principle}}
\end{figure}

Figure~\ref{fig:principle} gathers these ideas. Figure \ref{fig:principle}(a)
illustrates the physical picture for an incident SH2 mode scattering
into SH5. For this pair, the two depth orders p=3 and p=7 form the
kernel shown in Figure \ref{fig:principle}(b). Figure \ref{fig:principle}(c)
shows the resulting axial-depth spatial-frequency sampling when all
SH0-SH11 mode pairs are available at 2.0 MHz; restricting the data
to same- and adjacent-mode pairs leaves the much sparser coverage
in Figure \ref{fig:principle}(d). The comparison illustrates why
a sufficiently diverse set of cross-mode measurements is required
to separate the through-thickness orders. 

The transmission geometry also changes the relationship between axial
and depth spatial frequency compared with fundamental-mode reflection.
At 2.0 MHz in the 10 mm plate, for example, SH4-to-SH6 contributes
the sum order $p=10$ at $|K_{x}|=0.272\,\mathrm{mm}^{-1}$. In the
fundamental-mode reflection geometry demonstrated by Bonnetier et
al. \cite{bonnetierSmallDefectsReconstruction2022}, depth order 10
requires a propagating received mode of that order and therefore cannot
be sampled below $|K_{x}|=10\pi/h=3.14\,\mathrm{mm}^{-1}$. Thus a
high depth order does not require a large axial wavenumber: two high-order
transmitted modes can have a large sum order while their axial wavenumbers
remain close. This is directly relevant to material changes that vary
slowly along the plate but are confined through its thickness.

Expand the unknown contrast using the same unnormalised cosine basis,
\begin{equation}
q(x,z)=\sum_{p=0}^{P-1}a_{p}(x)\cos\left(\frac{p\pi z}{h}\right).\label{eq:qexpansion}
\end{equation}
The squared norm of the depth basis is $\chi_{0}=h$ and $\chi_{p}=h/2$
for $p>0$. Substituting the image expansion in Equation \eqref{eq:qexpansion}
and the two-order sensitivity in Equation \eqref{eq:twoorders} into
the calibrated measurement Equation \eqref{eq:canonicalmeasurement}
separates the axial and through-thickness integrals. Orthogonality
of the cosine basis makes the through-thickness integral zero for
every other depth order, leaving
\begin{equation}
d_{nm}=\sum_{p}\Gamma_{nmp}\int a_{p}(x)e^{\mathrm{i}K_{x}x}\,\mathrm{d}x,\label{eq:continuousforward}
\end{equation}
with
\begin{equation}
\Gamma_{nmp}=\kappa_{nm}\frac{N_{m}N_{n}}{2}\chi_{p}\left[(k_{m}k_{n}+\alpha_{m}\alpha_{n})\delta_{p,|m-n|}+(k_{m}k_{n}-\alpha_{m}\alpha_{n})\delta_{p,m+n}\right].\label{eq:gamma}
\end{equation}

If $|m-n|=m+n$, which occurs when either mode index is zero, the
two terms refer to the same depth order and are added. For measurement
row $j$, containing a particular $(m,n,\omega)$, \eqref{eq:continuousforward}
is written compactly as
\begin{equation}
d_{j}=\sum_{p=0}^{P-1}\Gamma_{jp}\int a_{p}(x)e^{\mathrm{i}K_{x,j}x}\,\mathrm{d}x.\label{eq:rowforward}
\end{equation}
Thus each mode-pair measurement contributes one complex equation involving
at most two depth orders at one generally non-uniform axial spatial
frequency. A single row cannot normally separate those depth orders;
separation is obtained by solving for all measurements together.

\subsection{Common measurement representation and axial discretisation}

Dataset-specific processing is completed before inversion. Each usable
measurement row contains the calibrated complex mode-pair measurement
$d_{j}$, its exact spatial-frequency coordinate $K_{x,j}$, the sparse
calibrated couplings $\Gamma_{jp}$, and a non-negative weight $w_{j}$.
Background referencing, mode identification, propagation correction
and conversion to the common Fourier convention are therefore isolated
from the reconstruction itself. The weights are fixed before inversion
and are target-independent: all usable ideal modal measurements are
given equal weight, while the practical datasets are weighted only
from background-calibration quality or uncertainty.

The reconstruction interval is divided into $L$ cells of width $\Delta x$
centred on $x_{l}$, and $a_{p}(x)$ is taken as constant within each
cell. The Fourier integral over one finite cell is then exact:
\begin{equation}
\int_{x_{l}-\Delta x/2}^{x_{l}+\Delta x/2}e^{\mathrm{i}K_{x}x}\,\mathrm{d}x=\Delta x\,\mathrm{sinc}\left(\frac{K_{x}\Delta x}{2}\right)e^{\mathrm{i}K_{x}x_{l}},\qquad\mathrm{sinc}(y)=\frac{\sin y}{y}.\label{eq:cell}
\end{equation}
The sinc term is simply the continuous Fourier transform of the rectangular
reconstruction cell. It is close to unity on a sufficiently fine grid,
but retaining it avoids a midpoint approximation at essentially no
additional computational cost. Defining
\[
s_{j}=\Delta x\,\mathrm{sinc}\left(\frac{K_{x,j}\Delta x}{2}\right),
\]
the discrete forward operator $A$ is
\begin{equation}
(Aa)_{j}=s_{j}\sum_{p=0}^{P-1}\Gamma_{jp}\sum_{l=0}^{L-1}a_{p,l}e^{\mathrm{i}K_{x,j}x_{l}}.\label{eq:discreteforward}
\end{equation}
The operator is evaluated matrix-free in chunks of measurement rows,
so the full $J\times PL$ complex matrix is never assembled.

\subsection{Regularised inverse and adjoint\label{sec:inverse}}

The physical contrast is real, although the measurements and operator
are complex. The coefficients are therefore obtained from the weighted,
regularised problem
\begin{equation}
\min_{a\in\mathbb{R}^{PL}}\left\Vert W^{1/2}(Aa-d)\right\Vert _{2}^{2}+\lambda^{2}\|La\|_{2}^{2},\qquad W=\mathrm{diag}(w_{j}).\label{eq:inverse}
\end{equation}
The regulariser approximates the first-order or $H^{1}$ seminorm,
\begin{equation}
\|La\|_{2}^{2}\simeq\int\!\!\int\left[\left(\frac{\partial q}{\partial x}\right)^{2}+\left(\frac{\partial q}{\partial z}\right)^{2}\right]\mathrm{d}z\,\mathrm{d}x.
\end{equation}
Axial variation is penalised using first differences of each $a_{p}$.
The depth term follows analytically from differentiating the cosine
basis, so higher depth orders receive progressively greater penalty
and $p=0$ has no depth penalty. The complete coefficient-space form
is given in Appendix \ref{sec:appendix-adjoint}.

The adjoint is defined using the complex inner product $\langle u,v\rangle=u^{H}v$.
Applying this definition to \eqref{eq:discreteforward} gives
\begin{equation}
(A^{H}y)_{p,l}=\sum_{j}\Gamma_{jp}^{*}s_{j}y_{j}e^{-\mathrm{i}K_{x,j}x_{l}}.\label{eq:adjoint}
\end{equation}
The conjugation of $\Gamma$ and reversal of the Fourier sign are
both required.

A real least-squares solver is used by stacking the real and imaginary
data equations:
\begin{equation}
\begin{bmatrix}\mathrm{Re}(W^{1/2}A)\\
\mathrm{Im}(W^{1/2}A)\\
\lambda L
\end{bmatrix}a\simeq\begin{bmatrix}\mathrm{Re}(W^{1/2}d)\\
\mathrm{Im}(W^{1/2}d)\\
0
\end{bmatrix}.\label{eq:realaugmented}
\end{equation}
Equation \eqref{eq:realaugmented} is solved using LSQR. LSQR is a
Krylov least-squares method that repeatedly applies the forward operator
and its adjoint rather than forming $A^{H}A$. The adjoint maps the
current measurement-space residual back into coefficient space and
therefore provides the information from which the solver constructs
its search directions; LSQR combines these directions rather than
simply taking steepest-descent steps. A single global phase offset
between the calibrated data and the forward-model convention is treated
as a nuisance parameter. In the LSQR reconstructions, starting from
zero phase, the data are rotated, Equation (21) is solved for the
real image coefficients, and the common phase is updated from the
argument of the complex inner product between the unphased prediction
and the measurements; three phase updates are used. This accounts
for a common phase-reference offset without altering the relative
modal phases. The explicit regularisation parameter $\lambda$ controls
the inverse. Iteration count is not used as an additional regularisation
parameter: the solver is run to its numerical stopping criterion,
with a maximum iteration count used only as a safeguard. 

After inversion, the depth-resolved image is synthesised directly
from the recovered coefficients,
\begin{equation}
q(x_{l},z_{i})=\sum_{p=0}^{P-1}a_{p,l}\cos\left(\frac{p\pi z_{i}}{h}\right).\label{eq:synthesis}
\end{equation}
Same-mode measurements constrain the $p=0$ component only through
its global axial integral; cross-mode-only inversions omit $p=0$
entirely. The latter images are therefore relative to their local
through-thickness mean; an independent local through-thickness measurement
could provide that missing mean.

\subsection{Depth information content}

The information available for depth reconstruction follows directly
from the modal coupling in Eq.~\eqref{eq:gamma}. Each measurement
contains at most two depth orders, while the set of propagating modes
determines which depth orders can be represented. If modes up to order
$M$ are available, the sum term can contain orders up to $p=2M$.
This is an upper limit rather than a guarantee of recovery: the relevant
depth orders must also be represented by sufficiently independent
mixtures across mode pairs and frequencies. As discussed above, the
$p=0$ component is not locally observable in this transmission geometry;
the depth-resolution question therefore concerns recovery of the non-zero
depth orders.

\subsection{Single-surface synthesis and calibration\label{sec:surface-synthesis}}

The ideal modal formulation above assumes direct access to individual
incident and received SH modes. For the practical scenario where access
is limited to a single surface, each physical excitation is instead
applied independently at one position on the accessible top surface,
$z=0$. In the two-dimensional $x$-$z$ model this is a $y$-directed
point source; physically, it represents a line source invariant in
$y$. An individual surface excitation launches a mixture of the propagating
SH modes. The required incident mode is formed by superposing responses
from several source positions so that a selected SH mode is synthesised
at a plane $x=x_{\mathrm{e}}$ at the edge of the imaging region.
The superposition is carried out in post-processing, exploiting linearity,
so simultaneous multi-element actuation is not required.

At a given frequency, the field at the synthesis plane due to source
$j$ is written as
\begin{equation}
u_{j}(x_{\mathrm{e}},z,\omega)=\sum_{m\in\mathcal{P}}G_{mj}(\omega)\phi_{m}(z),\label{eq:surface-source-field}
\end{equation}
where $\mathcal{P}$ is the set of propagating modes and $G_{mj}$
is the source-to-mode coefficient. For the relative synthesis used
here, the modal Green-function matrix is
\begin{equation}
G_{mj}(\omega)=\frac{\phi_{m}(0)}{2\mathrm{i}k_{m}}\exp\!\left[\mathrm{i}k_{m}(x_{\mathrm{e}}-x_{j})\right],\label{eq:surface-green}
\end{equation}
Here $\phi_{m}(0)$ is the value of the normalised mode shape at the
accessible surface and therefore sets the relative coupling of the
surface force to mode $m$. The factor $1/(2\mathrm{i}k_{m})$ and
the exponential are the corresponding modal Green-function amplitude
and propagation-phase terms. Any overall source scaling common to
all modes and source positions only sets the absolute field amplitude
and therefore does not affect the relative source weights.

Let $\boldsymbol{\beta}^{(q)}$ denote the vector of complex source
weights used to synthesise incident mode $q$. The modal-amplitude
vector at the synthesis plane is $G\boldsymbol{\beta}^{(q)}$, so
the weights are obtained directly from
\begin{equation}
G\boldsymbol{\beta}^{(q)}=\mathbf{e}_{q},\qquad\boldsymbol{\beta}^{(q)}=G^{\dagger}\mathbf{e}_{q}.\label{eq:surface-weights}
\end{equation}
Here $\mathbf{e}_{q}$ is the unit vector for the requested mode and
$^{\dagger}$ denotes the Moore-Penrose pseudoinverse. When there
are more source positions than retained propagating modes, this gives
the minimum-norm set of source amplitudes and phases. If $G$ has
full row rank, the same solution can be written
\begin{equation}
\boldsymbol{\beta}^{(q)}=G^{H}(GG^{H})^{-1}\mathbf{e}_{q}.\label{eq:surface-weights-fullrow}
\end{equation}
Thus the source weights are fixed by the reference-plate modal properties,
the selected source positions, the synthesis plane and the requested
incident mode; they are not fitted using the target response.

The synthesis equations require the modal wavenumbers $k_{m}(\omega)$,
but these need not be assumed to equal their nominal analytical values.
They are estimated from a homogeneous reference acquisition in which
each transmitter position is excited independently and the response
is recorded at multiple receiver positions beyond the opposite axial
edge of the imaging region. No mode-pure excitation is required for
this calibration: the raw point-source transfer data already contain
the spatial phase signatures of the propagating modes.

For a homogeneous plate, the reference response depends on the source-receiver
separation and can be represented at each frequency as
\begin{equation}
H_{rj}^{(0)}(\omega)\simeq\sum_{m\in\mathcal{P}}A_{m}(\omega)\exp\!\left[\mathrm{i}k_{m}(\omega)(x_{r}-x_{j})\right],\label{eq:reference-wavenumber-fit}
\end{equation}
where $H_{rj}^{(0)}$ is the homogeneous-reference response from transmitter
$j$ to receiver $r$, and $A_{m}$ is a complex coefficient containing
the modal excitation, reception and common phase/amplitude factors.
The model is nonlinear in the real wavenumbers $k_{m}$ but linear
in the complex coefficients $A_{m}$. For any trial set of wavenumbers,
the corresponding $A_{m}$ are therefore obtained by complex linear
least squares; an outer fit adjusts only the $k_{m}$ to minimise
the residual over the available source-receiver separations. The nominal
dispersion relation identifies the modal branches and supplies the
initial wavenumbers, while the multiple transmitter-receiver combinations
make the fitted reference wavenumbers overdetermined by the measured
data. These fitted wavenumbers are then used to evaluate $G$ in Eqs.~\eqref{eq:surface-green}-\eqref{eq:surface-weights-fullrow}
and the steering matrices used for modal separation. Thus the reference
wavenumbers are determined from the raw surface measurements before
the post-processed modal synthesis is formed.

The source positions themselves need not be uniformly spaced, and
in fact non-uniformly spaced positions can be selected to improve
the calibration process. Since the physical positions must be chosen
before the homogeneous reference data are acquired, their design uses
the nominal modal wavenumbers from the reference dispersion relation,
whereas the fitted reference wavenumbers are used for the modal synthesis
and separation. Their purpose is to make the modal phase patterns
across the aperture sufficiently independent that the synthesis and
modal separation remain well conditioned. Because the other factors
in $G_{mj}$ are independent of source position, position selection
can be based on the phase-only steering matrix $V$, with entries
\begin{equation}
V_{jm}=\exp[\mathrm{i}k_{m}(x_{j}-\bar{x})].
\end{equation}
For position selection, this matrix is evaluated using the nominal
$k_{m}$ values. Writing $\bar{x}$ as the aperture reference position,
Equation~\eqref{eq:surface-green} can be factorised as
\begin{equation}
G=\Lambda V^{H},\qquad\Lambda_{mm}=\frac{\phi_{m}(0)}{2\mathrm{i}k_{m}}\exp\!\left[\mathrm{i}k_{m}(x_{\mathrm{e}}-\bar{x})\right],\qquad GG^{H}=\Lambda(V^{H}V)\Lambda^{H}.\label{eq:surface-factorisation}
\end{equation}
The diagonal matrix $\Lambda$ contains the mode-dependent surface
coupling and propagation factors but is independent of which source
positions are selected. The source geometry therefore enters the conditioning
only through $V$. The modal phase-signature Gram matrix is $V^{H}V$;
it becomes poorly conditioned when two modal phase patterns are nearly
linearly dependent. A D-optimal design therefore selects positions
to maximise the log-determinant of this Gram matrix over the design
frequencies, improving modal separability and reducing noise amplification
\cite{fedorovOptimalExperimentalDesign2010}.

By reciprocity, the coupling coefficient from a surface position into
a guided mode is the same as the coefficient for that mode arriving
reciprocally at the same position, under the same modal normalisation.
An outgoing mode from the imaging region therefore produces across
the receiver aperture the same type of spatial phase signature represented
by the steering matrix. Projecting the measured receiver-position
vector onto these fitted signatures separates the received field into
modal amplitudes. Transmission and reception can consequently both
be handled as linear complex combinations of independently acquired
surface responses.

For a complete surface acquisition, let $H_{rj}$ be the measured
response at receiver position $r$ to a unit excitation at transmitter
position $j$, so that $H$ is the complete surface transfer matrix.
Let $V_{t}$ and $V_{r}$ be the phase-only steering matrices for
the transmitter and receiver apertures, evaluated using the fitted
homogeneous-reference wavenumbers. In this representation, if $N_{m}$
modes are retained, $H$ is an $N_{r}\times N_{t}$ matrix of physical
receiver-by-transmitter measurements, while $V_{t}$ and $V_{r}$
have sizes $N_{t}\times N_{m}$ and $N_{r}\times N_{m}$, respectively.
The remaining $N_{m}\times N_{m}$ matrix $C$ is the modal transfer
matrix: its element $C_{nm}$ is the complex transfer from incident
mode $m$ to received mode $n$. The steering matrices contain only
the spatial phase signatures of the modes; mode-dependent source and
receiver coupling, homogeneous propagation, and any scattering therefore
remain in $C$. With $C$ denoting the modal transfer matrix, the
surface data have the matrix form
\begin{equation}
H\simeq V_{r}CV_{t}^{H}.\label{eq:surface-matrix-model}
\end{equation}
Equation~\eqref{eq:surface-matrix-model} can be read from right
to left. For a physical transmitter excitation, $V_{t}^{H}$ supplies
the phase factors associated with each incident modal channel; $C$
transfers those incident channels into received modal channels; and
$V_{r}$ maps the received modal channels to the physical receiver
positions. Thus Equation~\eqref{eq:surface-matrix-model} is simply
a factorisation of the measured point-to-point surface transfer matrix
into transmitter phase patterns, modal transfer, and receiver phase
patterns. The least-squares modal separation is therefore
\begin{equation}
C=V_{r}^{\dagger}H(V_{t}^{H})^{\dagger},\label{eq:surface-modal-transfer}
\end{equation}
Equation~\eqref{eq:surface-modal-transfer} performs the inverse
change of coordinates in a least-squares sense. Left multiplication
by $V_{r}^{\dagger}$ combines the receiver-position measurements
to isolate each received modal phase pattern, while right multiplication
by $(V_{t}^{H})^{\dagger}$ performs the corresponding separation
over the independently excited transmitter positions. The complete
set of point-to-point measurements is thereby reduced to $C$: rows
of $C$ correspond to received modes and columns to incident modes.
The phase-only projection separates modal components using their fitted
spatial phase patterns, but it does not by itself remove the mode-dependent
source and receiver amplitudes or the remaining nominal propagation
phase. These factors therefore remain in the modal transfer matrix
$C$. Applying the same decomposition to homogeneous-reference and
perturbed datasets gives $C_{0}$ and $C_{1}$. For the homogeneous
plate, $C_{0}$ should be close to diagonal because there is no physical
conversion between distinct SH modes. Its diagonal element $(C_{0})_{mm}$
is the measured complex through-transmission for incident mode $m$,
including the source and receiver coupling and the homogeneous propagation
amplitude and phase. The corresponding matrix $C_{1}$ contains the
same baseline channel factors together with any perturbation-induced
mode conversion. The perturbed modal transfer is calibrated against
this reference as
\begin{equation}
M=C_{1}C_{0}^{\dagger}.\label{eq:surface-calibration}
\end{equation}
Equation~\eqref{eq:surface-calibration} acts on the right because
columns of $C$ correspond to incident modes. If the homogeneous matrix
were exactly diagonal, the operation would reduce to $(M)_{nm}\simeq(C_{1})_{nm}/(C_{0})_{mm}$:
every element in column $m$ of the perturbed modal transfer is normalised
by the measured homogeneous through-transmission of that same incident
mode. Using the full pseudoinverse $C_{0}^{\dagger}$ performs the
corresponding matrix de-embedding when finite-aperture modal separation
leaves small residual off-diagonal terms in $C_{0}$. If the perturbed
response is identical to the reference, this gives $M\simeq I$. An
off-diagonal element $M_{nm}$ therefore represents target-induced
conversion from incident mode $m$ to received mode $n$, relative
to the homogeneous transmission of the incident channel. This removes
the incident-channel source scaling and homogeneous through-propagation.
The remaining known relative receive-mode coupling is included in
the surface/modal normalisation used to convert $M_{nm}$ to the canonical
Born measurement. No target-minus-reference time-history subtraction
is required. For each retained off-diagonal mode pair, the corresponding
entry of $M$ is converted to the canonical scattering-data convention
of Equation~\eqref{eq:canonicalmeasurement} and supplies one measurement
$d_{j}$ in Equation~\eqref{eq:rowforward}. The matching $\Gamma_{jp}$
is evaluated from Equation~\eqref{eq:gamma} using the same fitted
modal wavenumbers and surface/modal normalisation. No target-dependent
amplitude scaling is applied.

\subsection{Evaluation metrics}

For compact targets, the reconstructed position is taken as the most
negative image value within a 2.0 mm-radius local search window around
its known centre. The localisation error is the Euclidean distance
in the $x$-$z$ plane between this reconstructed position and the
true target centre. Whole-image correlation is the Pearson correlation
coefficient between corresponding pixel values over the full displayed
$x$-$z$ reconstruction field.

OpenAI ChatGPT (GPT-5.6 Sol) was used during development of the numerical
workflow to assist with code review and debugging. The numerical algorithms
were specified by the author, and all AI-assisted implementation changes
and outputs were independently checked using the analytical comparisons,
adjoint tests, matched-model reconstructions and finite-element/Born
comparisons described in this paper.

\section{Two-dimensional evaluation\label{sec:2d-evaluation}}

The two-dimensional studies use a 10 mm thick isotropic reference
plate with Young's modulus 210 GPa, Poisson's ratio 0.3125 and density
8000 kg m\textsuperscript{-3}, corresponding to a shear-wave speed
of approximately 3162 m/s. Changes are specified as fractional shear-modulus
perturbations. Unless explicitly identified as a matched-model reference
reconstruction, all numerical wavefield datasets are generated using
Pogo~\cite{huthwaiteAcceleratedFiniteElement2014} finite-element
simulations.

\subsection{Ideal modal model}

The ideal case uses a 180 mm long two-dimensional SH model with 0.1
mm \texttimes{} 0.1 mm SH4R element size. A 20-cycle Hann-windowed
excitation is centred at 2.0 MHz and excites modes SH0 to SH12. SH0
to SH11 are retained for reconstruction; SH12 is excluded because
it lies too close to cut-off over the lower part of the analysed frequency
band. Modal scattering data are analysed from 1.91 to 2.04 MHz. Each
incident SH mode is excited in a separate finite-element shot by applying
its analytical through-thickness mode shape as a distributed force
at an internal source plane; received modal amplitudes are obtained
by projection onto the corresponding analytical mode shapes at receiver
planes. This is an idealised configuration because access through
the thickness of the plate is generally not possible, and is used
as an initial assessment of the method before considering more practical
surface-based transduction.

Four circular perturbations of radius 0.75 mm are centred at x = -12,
-4, 4 and 12 mm and depths of 1.5, 3.7, 6.3 and 8.5 mm, with shear-modulus
changes of -2\%, -4\%, -6\% and -8\%, respectively. The common inverse
is supplied with calibrated modal scattering data extracted from the
finite-element histories for this configuration. In this ideal modal
case, the perturbed and homogeneous modal histories are subtracted
and then normalised by the incident-mode amplitude measured in the
homogeneous model. The dataset includes same-mode measurements, which
constrain the global axial integral of the depth-average order but
do not localise that mean along the plate.

The inverse uses a reconstruction interval from -20 to +20 mm with
0.25 mm cells and depth orders $p=0,\ldots,22$. For the two-dimensional
evaluations, the regularisation strength is scaled to the data and
regularisation operators according to
\begin{equation}
\lambda=3\,\frac{\mathrm{median}_{j:\|W^{1/2}A_{:j}\|_{2}>0}\|W^{1/2}A_{:j}\|_{2}}{\mathrm{median}_{j:\|L_{:j}\|_{2}>0}\|L_{:j}\|_{2}}.\label{eq:lambda-2d}
\end{equation}
The prefactor 3 is fixed from the ideal validation so that the phase-corrected
residual remains commensurate with the independently observed approximately
6\% discrepancy between the Born prediction and the finite-element
scattering data, rather than forcing the inverse to fit below the
forward-model error. The same prefactor is then held fixed for the
single-surface, noise and thickness-mismatch cases. For the present
ideal dataset this gives $\lambda=2.26911$. LSQR uses absolute and
residual tolerances of $10^{-7}$ and a maximum of 350 iterations.
No target geometry or reconstructed image is used to tune $\lambda$.

\begin{figure}
\includegraphics[width=1\textwidth]{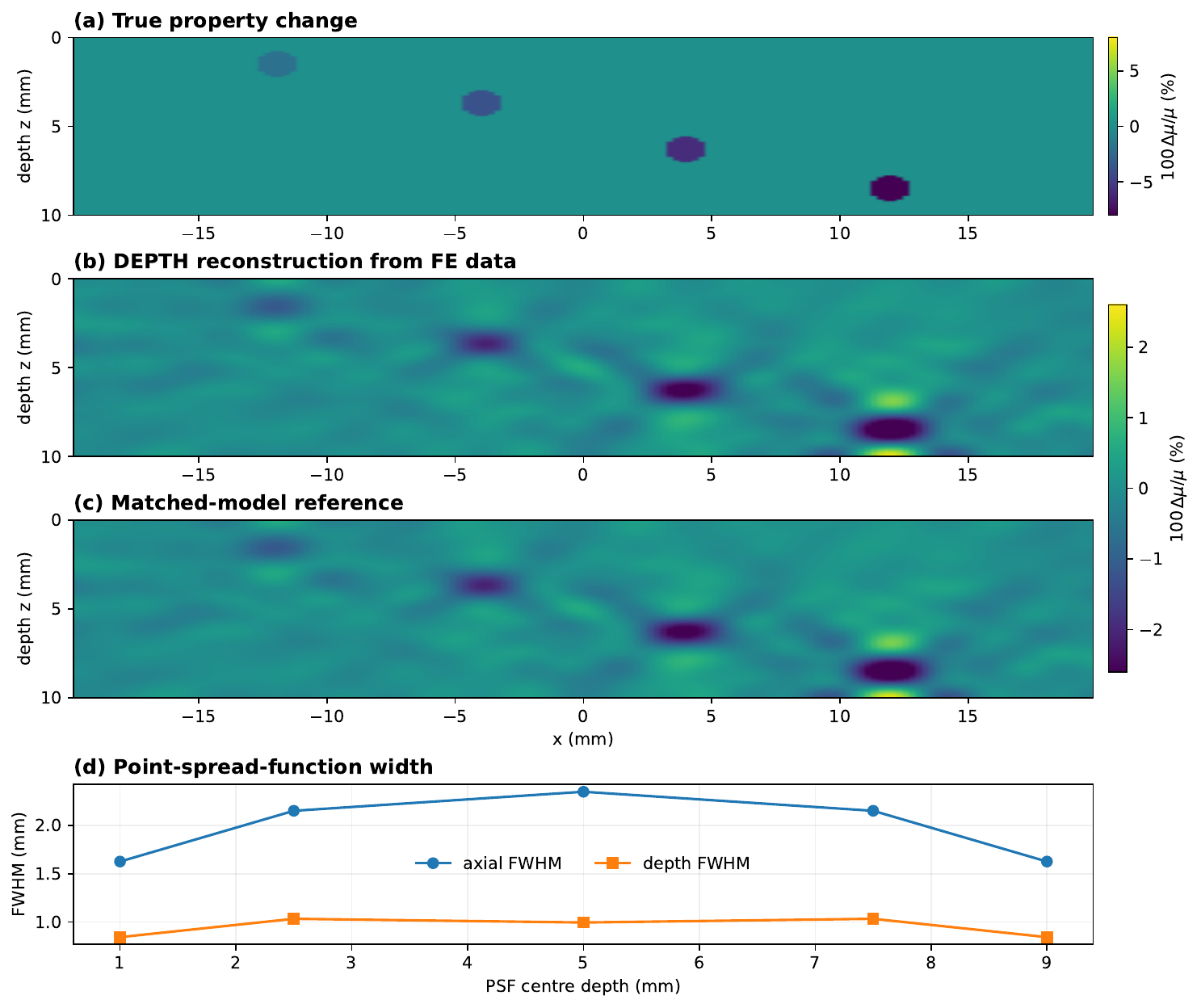}\caption{Ideal two-dimensional DEPTH reconstruction and resolution. (a) Applied
fractional shear-modulus perturbation. (b) Reconstruction from calibrated
modal scattering data extracted from the finite-element simulations.
(c) Matched-model reference reconstruction obtained by generating
data from the known phantom with the same linear forward model used
by the inverse and applying the same regularised inverse. (d) Axial
and through-thickness FWHM of the point-spread function versus depth,
used here as a compact measure of the main response width rather than
a complete definition of resolution. The agreement between (b) and
(c) shows that the main loss of peak contrast and surrounding structure
are predicted by the finite resolution of the inverse.\label{fig:ideal-2d-validation}}
\end{figure}

Figure~\ref{fig:ideal-2d-validation}(a) shows the applied fractional
shear-modulus perturbations, while Figure \ref{fig:ideal-2d-validation}(b)
shows the reconstruction from the finite-element modal scattering
data. All four perturbations are recovered at the correct axial and
through-thickness positions. The largest localisation error is 0.125
mm in x and 0.10 mm in z, corresponding to half an axial image cell
and one displayed depth sample, respectively. The reconstructed peak
modulus changes are smaller than the literal target values, ranging
from -1.16\% for the -2\% target to -3.87\% for the -8\% target. As
discussed in Section~\ref{sec:inverse}, the ideal same-mode data
determine only the overall through-thickness mean, not how that mean
varies with axial position. The $p=0$ part of the reconstruction
is therefore a constant offset and does not affect the recovered target
locations. The final relative complex-data residual is 5.6\%. The
5.6\% value is the unweighted Euclidean norm of the complex data misfit,
after fitting the global phase, divided by the norm of the finite-element
data. The inverse contains 3680 real coefficients and 9648 complex
measurement rows. On an HP Z820 workstation with two 8-core Intel
Xeon E5-2665 processors at 2.40 GHz and 48 GB RAM, this inversion
required 128 LSQR iterations and 6.5 s.

Independently, a direct data-domain comparison over the SH0-SH11 transmission
channels gives a complex correlation of 0.999 between the calibrated
first-Born prediction and the finite-element scattering data, with
a 5.5\% residual after allowing a single global complex scale; for
cross-mode channels alone the corresponding values are 0.998 and 6.4\%.
These Born-data metrics use SH0-SH11 transmission channels with normalised
finite-element scattered amplitude above -50 dB; the quoted residual
is the unweighted Euclidean norm of the difference between the best-scaled
Born prediction and the finite-element data, divided by the Euclidean
norm of the finite-element data.

Figure \ref{fig:ideal-2d-validation}(c) shows the corresponding matched-model
reference reconstruction, obtained using data generated by the same
linear forward model used in the inversion. This is a deliberate implementation
of the so-called “inverse crime”, normally regarded as a trivial validation
case because the data-generation and inversion models are identical.
Here it is used instead as a reference defining what is theoretically
achievable from the given measurement setup and regularised inverse.
Comparison with the finite-element reconstruction therefore separates
limitations imposed by the acquisition and inversion from discrepancies
between the finite-element scattering and the linearised Born model.
To determine whether this contrast loss is caused by discrepancy between
the finite-element scattering and the linearised model, or by the
finite resolution of the inverse, the known phantom is passed through
the same discrete linear forward model used by the inverse and reconstructed
using exactly the same regularised inverse. The resulting reference
reconstruction has a whole-image correlation of 0.998 with the finite-element
reconstruction. Its target peaks are -1.17\%, -2.12\%, -2.98\% and
-3.76\%, compared with -1.16\%, -2.12\%, -3.05\% and -3.87\% from
the finite-element data. The common colour limits used for the reconstruction
figures are chosen to make spatial structure comparable and therefore
saturate some negative extrema. The dominant loss of peak contrast
and the surrounding oscillatory structure are therefore set by the
measurement and inversion resolution rather than by discrepancy between
the finite-element scattering and the Born model. Because the phantom
spans contrasts from -2\% to -8\%, this agreement provides an end-to-end
check that the first-order linearisation is adequate over the contrast
range and inclusion size used here; it should not be interpreted as
a general Born-validity limit for larger or differently shaped perturbations.

Figure \ref{fig:ideal-2d-validation}(d) gives a complementary measure
of the point-spread width independently of the finite targets. The
point-spread functions are generated entirely with the matched linear
forward and inverse operators, without finite-element data. A unit
point-like perturbation represented in the same finite axial and cosine-depth
basis is placed at five depths, passed through the same regularised
forward and inverse operators, and the FWHM is measured from axial
and through-thickness cuts through each reconstructed point-spread
function. FWHM is used here as a compact measure of the width of the
main point-spread response rather than as a complete definition of
resolution. Across depths of 1-9 mm, the through-thickness FWHM is
0.84-1.03 mm and the axial FWHM 1.63-2.35 mm. Thus, for this SH0-SH11
measurement set, the characteristic through-thickness response width
remains close to 1 mm throughout the 10 mm thickness. This establishes
the key result of the ideal two-dimensional case: multimode SH scattering
contains sufficient independent information to resolve local changes
in shear modulus both along the plate and through its thickness.

\subsection{Single-surface transduction model}

The single-surface evaluation applies the synthesis and calibration
of Section~\ref{sec:surface-synthesis} to a 560 mm long two-dimensional
plate with separate transmitter and receiver apertures on the accessible
surface, each 150 mm long and containing 20 selected positions flanking
a 160 mm imaging region. Candidate locations are spaced by 0.60 mm;
the D-optimal selection uses five design frequencies from 1.5 to 1.9
MHz and a fixed $10^{-9}I$ ridge in the Gram matrix during the early
rank-deficient steps. The resulting 20 transmitter positions were
\textminus 244.98, \textminus 238.38, \textminus 231.18, \textminus 227.58,
\textminus 221.58, \textminus 216.18, \textminus 195.78, \textminus 178.98,
\textminus 172.98, \textminus 164.58, \textminus 156.78, \textminus 150.78,
\textminus 142.98, \textminus 137.58, \textminus 132.78, \textminus 129.78,
\textminus 121.38, \textminus 111.18, \textminus 100.38 and \textminus 94.98
mm, with the receiver positions mirrored about $x=0$.

The excitation is a 10-cycle Hann-windowed toneburst centred at 1.7
MHz, and modal transfer data are analysed from 1.5 to 1.9 MHz. The
model contains the same four circular perturbations as Figure~\ref{fig:ideal-2d-validation}.
Homogeneous-reference and perturbed datasets are generated using two-dimensional
Pogo finite-element simulations with the same mesh, source/receiver
positions and excitation, so the imposed material perturbation is
the only change between them.

The inverse uses a 160 mm axial reconstruction interval with 0.25
mm cells and depth orders $p=1,\ldots,21$. The regularisation strength
is set by Equation~\eqref{eq:lambda-2d}, giving $\lambda=1.103955$
for the nominal reconstruction. As before, LSQR uses absolute and
residual tolerances of $10^{-7}$ and a maximum of 350 iterations.

\begin{figure}
\includegraphics[width=1\textwidth]{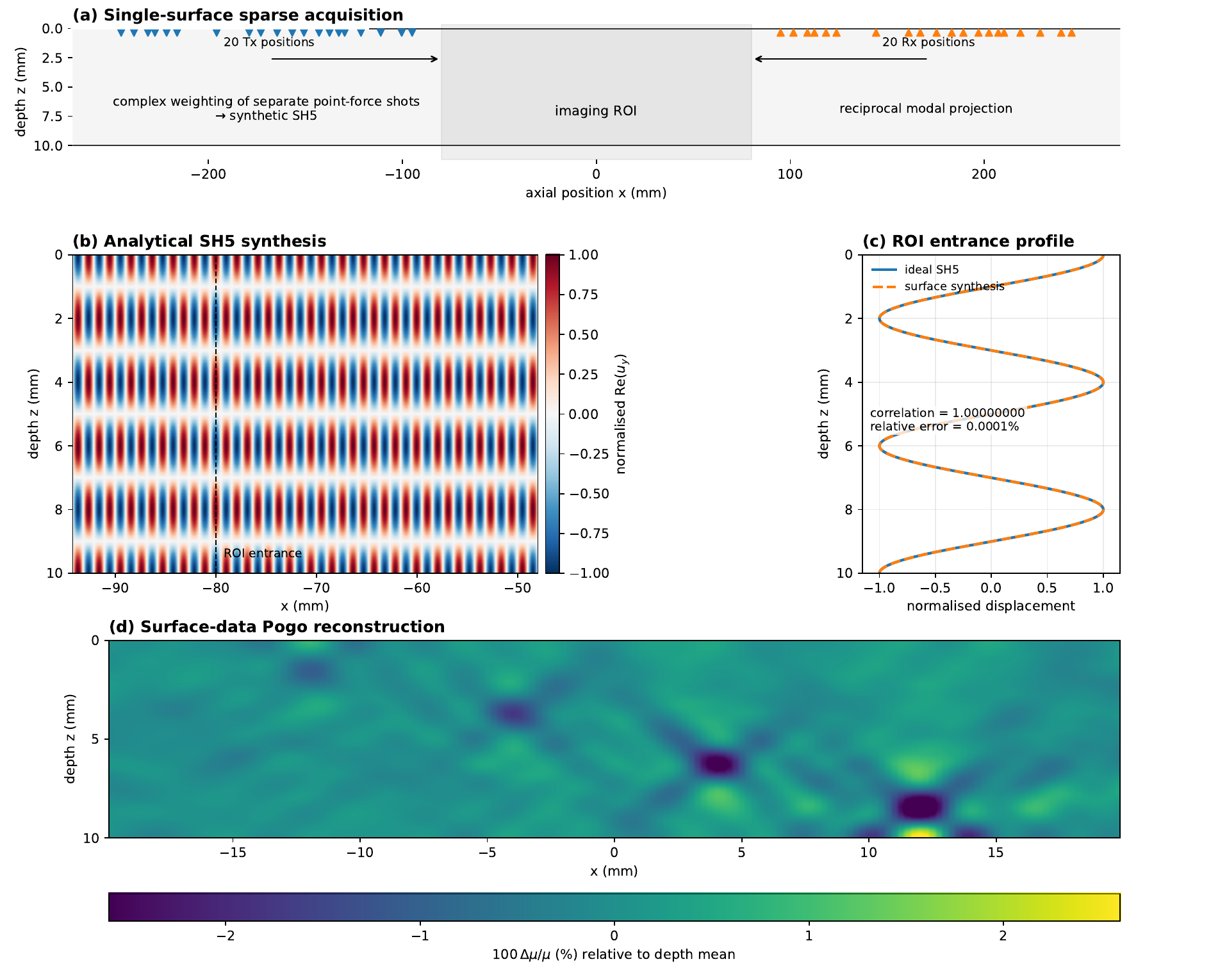}\caption{Single-surface access validation. (a) Sparse 20-position transmitter
and receiver apertures surrounding the imaging region. (b) Analytical
SH5 field synthesised at 1.70 MHz by complex weighting of the separate
surface point-force shots. (c) Through-thickness profile at the entrance
to the imaging region compared with the analytical SH5 mode. (d) $H^{1}$-regularised
DEPTH reconstruction from finite-element surface data for the same
four circular perturbations used in Figure~\ref{fig:ideal-2d-validation}.
The numerical colour limits in Figure 3(d) are identical to those
used for the ideal finite-element reconstruction in Figure~\ref{fig:ideal-2d-validation}(b);
because only cross-mode measurements are retained, $\Delta\mu/\mu$
is shown relative to its through-thickness mean.\label{fig:surface-validation}}
\end{figure}

Figure~\ref{fig:surface-validation}(a) shows the resulting single-surface
acquisition geometry. Figures~\ref{fig:surface-validation}(b) and
(c) verify the analytical synthesis construction for SH5 at 1.70 MHz:
at the entrance to the imaging region the synthesised through-thickness
profile matches the analytical SH5 profile to numerical precision.
The finite-element surface data are assessed separately through the
homogeneous modal fit, which also supplies the conditioning and frequency-selection
criteria.

Target-independent background calibration and conditioning criteria
retain 43 of the 81 analysed frequencies, spanning 1.50-1.88 MHz,
and yield 4882 complex cross-mode measurements. Frequencies are retained
only when the homogeneous modal-fit residual is below 0.05, the steering-matrix
condition number is below 8, the background modal-transfer condition
number is below 12 and the frequency does not exceed 1.88 MHz; propagating
modal channels additionally require the axial wavenumber to be at
least 15\% of the bulk wavenumber. Each row is weighted by the reciprocal
of one plus the fourth power of the ratio between its background modal-fit
residual and 0.05. Across the retained set, the median steering and
background-modal condition numbers are 4.9 and 5.9, respectively.
Figure~\ref{fig:surface-validation}(d) shows the reconstruction
from the finite-element surface data for the same four-target phantom
used in Figure~\ref{fig:ideal-2d-validation}. The principal result
is unchanged from the ideal modal case: all four targets are resolved
at the correct axial and through-thickness locations despite excitation
and measurement being restricted to one surface. As only cross-mode
measurements are retained in this configuration, the local $p=0$
depth-average component is unavailable and the plotted $\Delta\mu/\mu$
is therefore relative to its through-thickness mean. The recovered
local extrema for the four targets are -1.06\%, -1.77\%, -3.06\% and
-4.00\%, compared with -1.16\%, -2.12\%, -3.05\% and -3.87\% in the
ideal modal reconstruction of Figure~\ref{fig:ideal-2d-validation}(b).
The two deeper targets therefore retain essentially the same reconstructed
contrast, while the two shallower targets have lower recovered extrema.
Because the ideal and surface datasets differ in analysed band, retained
rows and treatment of the depth-average order, this contrast difference
is descriptive rather than attributable to surface access alone. A
matched-model reference reconstruction using the same surface measurement
set (not shown) has a whole-image correlation of 0.978 with the finite-element
reconstruction.

\section{Three-dimensional finite-element evaluation\label{sec:3d-evaluation}}

The two-dimensional studies above provide a direct test of the modal
formulation and allow the depth-imaging behaviour to be interpreted
cleanly, but they assume invariance in the transverse direction. In
practice, both the transducers and the material changes being imaged
have finite transverse extent, so the measured response represents
a weighted transverse slice rather than an exactly two-dimensional
field. A three-dimensional finite-element model is therefore used
to test whether the two-dimensional x-z DEPTH reconstruction remains
useful when these effects are present in the measured data.

\subsection{Model setup}

\begin{figure}
\includegraphics[width=1\textwidth]{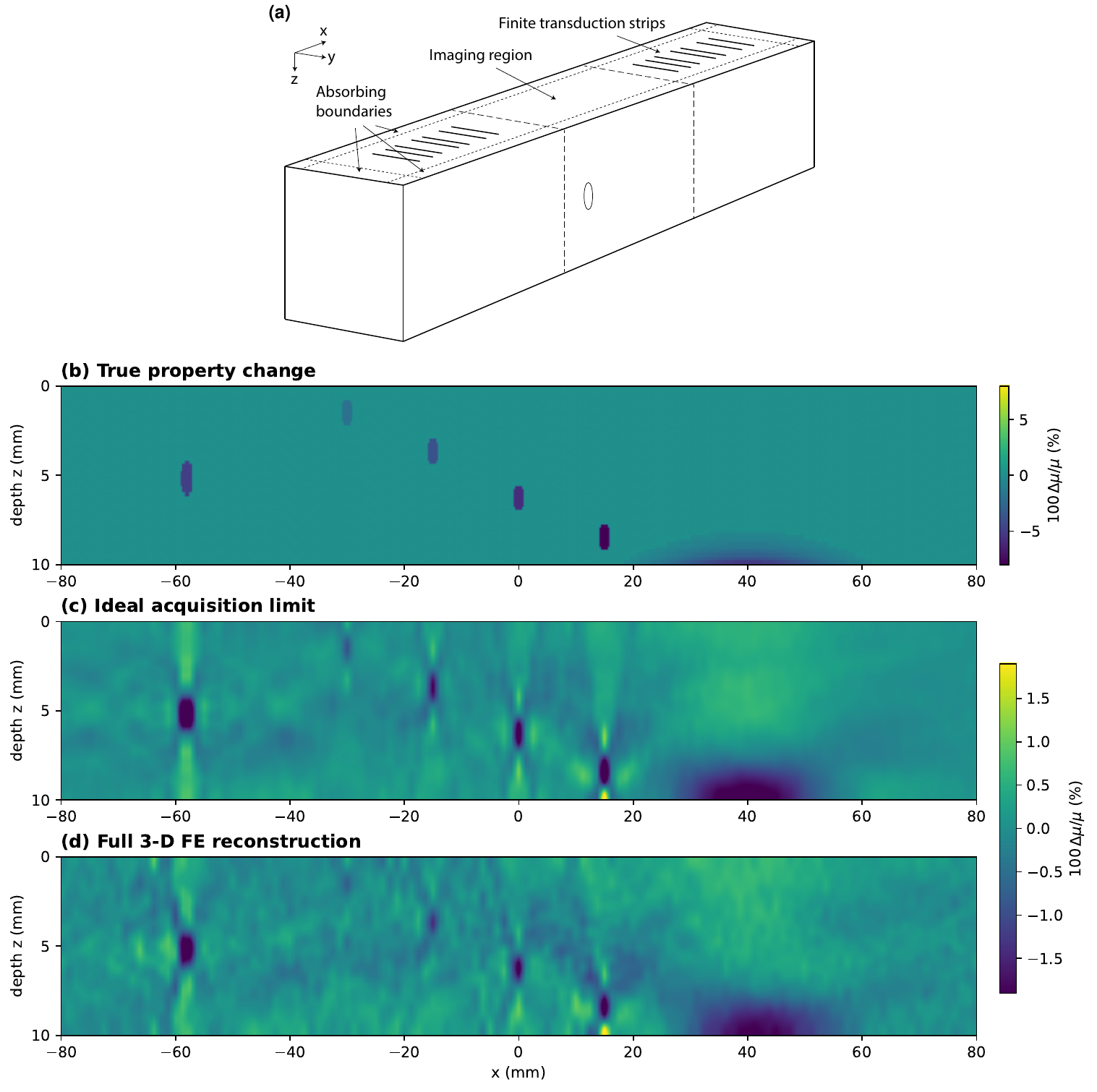}\caption{Three-dimensional finite-element validation. (a) Full three-dimensional
model geometry showing the finite transduction strips, imaging region
and absorbing boundaries. (b) Central $y=0$ section of the finite-element-applied
property change across the imaging area. (c) Matched acquisition-limited
ideal reconstruction using the same measurement rows, weighting and
regularisation as the finite-element case. (d) DEPTH reconstruction
from the full three-dimensional Pogo data. Reconstructions in (c)
and (d) use common numerical colour limits; because only cross-mode
measurements are retained, the reconstructed fields contain the depth-varying
component relative to its through-thickness mean.\label{fig:3d-validation}}
\end{figure}

Figure~\ref{fig:3d-validation}(a) illustrates the full three-dimensional
model geometry, including the finite transduction strips, imaging
region and absorbing boundaries. The three-dimensional validation
uses Pogo finite-element simulations of a 560 \texttimes{} 50 \texttimes{}
10 mm plate with 0.20 \texttimes{} 0.25 \texttimes{} 0.20 mm spatial
discretisation, C3D8R reduced-integration hexahedral elements and
a 10 ns timestep. The mesh contains 28.0 million elements; 30 mm and
8 mm absorbing regions are used at the axial and transverse boundaries,
respectively, while the plate surfaces remain traction free. The acquisition
again contains 20 transmitter and 20 receiver positions, represented
by 20 mm wide transverse strips, with a 10-cycle 1.7 MHz excitation
analysed over 1.5 to 1.9 MHz. 

Four compact ellipsoidal targets are distributed through the plate
depth, together with a y-invariant control target and a finite lower-surface
patch with Gaussian in-plane variation and 0.75 mm exponential depth
decay. The ellipsoids have semi-axes of 0.90 mm axially, 10 mm transversely
and 0.75 mm through the thickness, giving a 20 mm full transverse
extent. The four finite ellipsoids are centred at axial positions
-30, -15, 0 and 15 mm and depths 1.5, 3.7, 6.3 and 8.5 mm, with shear-modulus
changes of -2\%, -4\%, -6\% and -8\%, respectively. The y-invariant
control is centred at (-58, 5.2) mm with a -5\% change, while the
lower-surface patch is centred at x = 40 mm, has Gaussian standard
deviations of 10 mm and 15 mm in the axial and transverse directions,
respectively, and a nominal peak -6\% change. This patch is intended
to be representative of early-stage near-backwall material damage,
such as that associated with fatigue damage. For the fatigue patch,
the continuous target profile is assigned at element centres using
31 discrete material levels, with changes smaller than 0.5\% in magnitude
set back to the background material.

Baseline and perturbed simulations are processed to obtain calibrated
modal transfer measurements before applying the same reconstruction
operator. The modal wavenumbers are estimated using a structured-grid
SH dispersion approximation that accounts for axial and through-thickness
spatial discretisation. Fixed mode-dependent correction factors, obtained
from an earlier homogeneous three-dimensional finite-element calibration
using the same mesh, material properties and timestep, are then held
fixed for the target calculation. The homogeneous simulation again
provides the transfer-matrix calibration rather than being subtracted
directly from the perturbed time histories. As for the surface model,
the retained rows are cross-mode measurements and do not contain the
local depth-average order. The inversion itself remains two-dimensional
in x and z; the purpose of this case is therefore to test whether
the reduced modal reconstruction remains valid when the measured data
include full three-dimensional propagation, finite-width transduction
and finite transverse target extent. 

The reconstruction uses coefficient positions from -80 to +80 mm at
1.25 mm spacing and depth orders $p=1,\ldots,21$. The diagonal data
weights approximate frequency- and mode-pair-dependent inverse variances
inferred from homogeneous-background calibration residuals and are
normalised to unit median; correlations between derived modal measurement
rows are neglected. Generalised cross-validation using these fixed
background-derived weights selects the regularisation strength without
reference to target truth. In the common unnormalised-cosine convention
of Equation~\eqref{eq:inverse} this gives $\lambda=3.11093$. LSQR
in this case uses absolute and residual tolerances of $10^{-7}$ and
a maximum of 500 iterations.

\subsection{Reconstruction performance}

All 81 analysed frequencies from 1.50 to 1.90 MHz contribute to the
reconstruction, giving 9402 complex cross-mode measurements. The corresponding
inverse contains 2709 real coefficients; its regularisation is selected
before any comparison with target truth. Starting from the precomputed
frequency-domain surface-transfer data, the complete processing script
for this case performs weighting, generalised cross-validation selection
of the regularisation strength, reconstruction and generation of the
validation outputs. 

Figure~\ref{fig:3d-validation}(a) summarises the three-dimensional
geometry, while Figure \ref{fig:3d-validation}(b) shows the central
$y=0$ section of the finite-element-applied property field. Figures
\ref{fig:3d-validation}(c) and \ref{fig:3d-validation}(d) show the
matched acquisition-limited reconstruction and the reconstruction
obtained from the full three-dimensional finite-element data, respectively.
The matched ideal reconstruction is generated from the finite-element-applied
target map using the same two-dimensional forward model, measurement
rows, weighting and regularisation used for the finite-element data.
It therefore shows the image expected if the two-dimensional model
were exact. The difference between Figure \ref{fig:3d-validation}(c)
and Figure \ref{fig:3d-validation}(d) then shows the additional effect
of full three-dimensional propagation, finite-width transduction and
finite transverse target extent.

The four finite ellipsoidal targets are all recovered at the correct
axial positions and through-thickness depths. Using the target-location
definition introduced in the two-dimensional case, their x-z localisation
errors are 0, 0.050, 0.050 and 0.083 mm, respectively. The full finite-element
reconstruction has a whole-image correlation of 0.899 with the matched
acquisition-limited reconstruction. The y-invariant control target
is recovered with a 0.50 mm localisation error; essentially the same
error occurs in the matched acquisition-limited reconstruction, indicating
that this offset is set by the acquisition and inverse resolution
rather than by three-dimensional model mismatch.

The finite lower-surface patch is also recovered attached to the correct
surface. The nominal -6\% surface contrast corresponds to a -5.20\%
peak in the elementwise finite-element material field after discretisation.
Its reconstructed surface centroid is x = 42.86 mm compared with a
true centre of 40 mm, and its axial FWHM is 20.43 mm compared with
24.00 mm for the discretised finite-element target. The reconstructed
through-thickness profile has an effective depth of 0.957 mm, compared
with 0.570 mm for the finite-element-applied target. The nominal continuous
target has an exponential decay length of 0.75 mm; element-centred
assignment, material quantisation and the 0.5\% contrast cutoff account
for the difference between this nominal parameter and the effective
depth of the applied finite-element target. This effective depth is
defined as the amplitude-weighted mean distance inward from the lower
surface, using only the negative part of the reconstructed centreline
contrast; this first-moment measure is used rather than fitting an
exponential to the band-limited reconstruction.

A simple aperture estimate gives the expected scale of the transverse
region contributing to this reconstruction. At 1.70 MHz the shear
wavelength is 1.86 mm. For a uniform strip of width $D=20\,\mathrm{mm}$,
the one-way complex field amplitude follows the usual sinc aperture
pattern. For comparable transmitter and receiver propagation distances,
the linear Born sensitivity is the product of the two field amplitudes
and is therefore approximately sinc-squared, with transverse FWHM
$0.886\lambda_{s}R/D$, where $R$ denotes the propagation distance.
At the centre of the imaging region, where the propagation distance
from either strip is approximately 170 mm, this gives a two-way linear-sensitivity
FWHM of about 14.0 mm. Accounting for the unequal transmitter and
receiver propagation distances over the axial range -30 to +40 mm
occupied by the finite targets, the same linear-sensitivity estimate
varies from approximately 12.8 to 14.0 mm. If the squared magnitude
of the sensitivity is used instead, the corresponding FWHM is approximately
9.2 to 10.1 mm; the linear complex-amplitude sensitivity is the convention
used here. This is only a geometric aperture estimate rather than
a full three-dimensional sensitivity calculation. However, the compact
ellipsoids have a 20 mm full transverse width and the lower-surface
patch has a transverse FWHM of approximately 35 mm, both wider than
the estimated measurement slice. The retention of the target positions
and depths in Figure~\ref{fig:3d-validation}(d) is therefore consistent
with the central x-z section remaining a useful description for targets
that vary more slowly across y than the measurement sensitivity. The
finite transverse extent does, however, reduce the recovered contrast.
This is particularly visible for the shallow -2\% ellipsoid, whose
reconstructed peak is -0.76\% in the three-dimensional finite-element
case compared with -1.08\% in the matched ideal reconstruction. The
present case therefore supports use of the two-dimensional inverse
for targets whose transverse extent is at least comparable with the
measurement slice; it does not establish accuracy for narrow or strongly
off-axis targets, for which a three-dimensional forward model or additional
transverse sampling would be required. The present treatment is deliberately
slice-based. A fully three-dimensional extension could combine measurements
from multiple transverse offsets in a synthetic-aperture reconstruction
to recover variation in $y$ rather than treating it through a weighted
transverse slice; this is beyond the scope of the present study.

\section{Sensitivity to measurement noise and thickness uncertainty\label{sec:sensitivity}}

The preceding evaluations are promising, but the practical single-surface
reconstruction asks a relatively sparse set of surface measurements
to separate several depth orders and assumes that the plate geometry
is known. It is therefore important to establish whether measurement
noise and thickness uncertainty produce a gradual loss of image quality
or an abrupt failure of the depth localisation. Measurement noise
and uncertainty in assumed wall thickness are tested explicitly. The
single-surface two-dimensional model is used for these tests because
it retains the surface excitation, modal calibration and projection
stages without the cost of repeated three-dimensional simulations.

\begin{figure}
\includegraphics[width=1\textwidth]{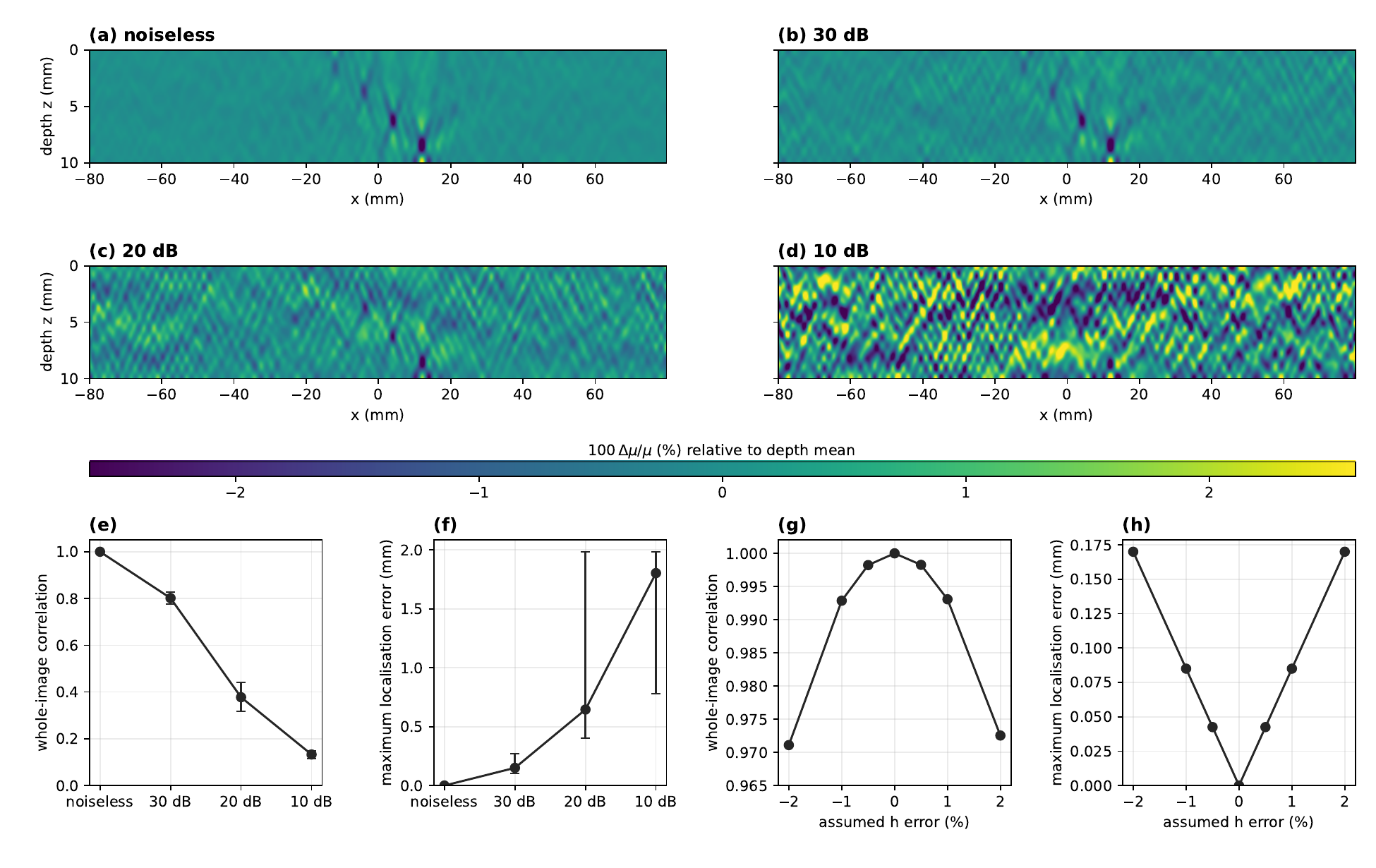}\caption{Sensitivity of the single-surface reconstruction to measurement noise
and assumed plate thickness. (a-d) Representative reconstructions
for the noiseless case and for 30, 20 and 10 dB additive circular
complex Gaussian noise, respectively. For each noisy case the displayed
reconstruction is the realisation whose whole-image correlation is
closest to the median over 20 independent realisations; all four reconstructions
use common numerical colour limits. (e) Median whole-image correlation
with the noiseless reconstruction and (f) median maximum target-localisation
error; error bars show the central 68\% interval (16th-84th percentiles)
over the 20 realisations. (g,h) Deterministic sensitivity to assumed
plate-thickness error with the empirically calibrated modal wavenumbers
held fixed, shown by whole-image correlation and maximum target-localisation
error, respectively.\label{fig:robustness}}
\end{figure}

For the noise study, the reconstruction grid, depth orders, LSQR tolerances
and regularisation prescription are the same as for the nominal single-surface
case. Additive circular complex Gaussian noise is applied to the complex
surface-transfer measurements before modal projection. Specifically,
independent zero-mean Gaussian noise is added to the real and imaginary
parts of every complex transmitter-receiver transfer value, with the
complex noise amplitude scaled so that its RMS over the 20 by 20 channels
at the 43 nominally retained frequencies gives the requested SNR.
The nominal modal calibration, geometry and source/receiver positions
are held fixed, so this isolates additive surface-measurement noise
rather than calibration, coupling or positioning drift. The signal-to-noise
ratio is defined as $20\log_{10}[\mathrm{rms}(|d|)/\mathrm{rms}(|n|)]$,
where $d$ is the noise-free complex surface-transfer data and $n$
is the added noise. The RMS is evaluated over all 20 by 20 surface-transfer
channels at the 43 frequencies retained by the nominal reconstruction.
Twenty independent realisations are reconstructed at 30, 20 and 10
dB. Figure~\ref{fig:robustness}(a-d) shows the noiseless reconstruction
and, at each noise level, the realisation whose whole-image correlation
is closest to the median across the 20 runs. This gives an objective
example that is representative according to the reported image metric
rather than selecting a visually favourable reconstruction. The same
numerical colour limits are used throughout. At 30 dB the target structure
is retained with only a modest increase in background artefacts. At
20 dB the targets remain visible in the representative reconstruction,
but the distributed background structure is substantially increased.
At 10 dB the reconstruction is dominated by noise-related structure.

The quantitative trends are shown in Figure~\ref{fig:robustness}(e,f).
All 20 realisations produce a reconstruction at each noise level.
The median whole-image correlation decreases from 0.802 at 30 dB to
0.378 at 20 dB and 0.133 at 10 dB. The corresponding median maximum
target-localisation errors are 0.150, 0.645 and 1.803 mm, respectively.
These values are bounded by the 2.0 mm-radius local search window
used by the localisation metric; the 1.803 mm median error at 10 dB
is therefore already close to the largest error that this metric can
report. Error bars show the central 68\% interval, taken as the 16th-84th
percentiles. This is analogous to a one-standard-deviation interval
for a Gaussian distribution, while making no Gaussian assumption about
the Monte Carlo samples. The spread in localisation performance becomes
substantial by 20 dB. At the highest noise levels the localisation
metric itself also becomes less informative: each target position
is defined by the most negative value within the prescribed local
search region around its known centre, so once the target response
is obscured the metric increasingly reports the strongest noise-related
fluctuation inside that region rather than unconstrained target localisation.

The second test isolates an assumed-thickness error in the depth model
after modal-dispersion calibration. The modal wavenumbers $k_{m}(\omega)$
estimated from the reference transfer response are held fixed, while
the assumed thickness $h$ used for the mode shapes, modal coupling,
depth basis and regularisation is varied from -2\% to +2\%, corresponding
to 9.8-10.2 mm about the true 10 mm value. This is deliberately narrower
than a complete mismatch between a calibration location and an inspection
location: it tests a geometry error in the inverse after the modal
dispersion has been established, rather than also changing the measured
dispersion itself. Figure~\ref{fig:robustness}(g,h) shows a smooth
degradation over this range. The whole-image correlation with the
correctly modelled reconstruction remains between 0.971 and 1.000,
while the maximum target-localisation error remains at or below 0.170
mm. At a 1\% thickness error the correlation is approximately 0.993
and the maximum localisation error is 0.085 mm. The reconstructed
target amplitudes also change only weakly across the sweep. The quoted
maximum localisation errors are dominated by the change of depth coordinate
itself: the deepest target at 8.5 mm shifts by 0.085 mm for a 1\%
thickness change and 0.170 mm for 2\%, exactly the corresponding geometric
rescaling, while the recovered coefficients and amplitudes change
only weakly. The \textpm 2\% range should therefore be read primarily
as a check that the recovered depth-varying coefficients remain stable
under this isolated coordinate/model perturbation, not as an independent
localisation tolerance or a general tolerance on wall-thickness variation.
If the thickness varies more strongly across the inspection region,
the local geometry must be measured or incorporated into the forward
model rather than represented by a single value of $h$.

A practical implementation will need to establish the modal calibration
for the particular transduction system and component geometry. Wall
thickness is not necessarily an uncontrolled quantity: a conventional
local thickness measurement could be used to supply $h$ independently.
The multimode SH dispersion also contains, in principle, separate
information about thickness and background shear-wave speed. From
$k_{m}^{2}=\omega^{2}/c_{s}^{2}-(m\pi/h)^{2}$, at fixed frequency
the intercept of $k_{m}^{2}$ against $m^{2}$ is governed by $c_{s}$,
whereas the dependence on mode order is governed by $h$. Resolving
several mode orders and frequencies therefore provides an overdetermined
route to estimate thickness and background shear speed jointly, or
an independent wall-thickness measurement can be used to constrain
that fit. This distinction is important because a dispersion change
caused by geometry should not be interpreted as a local modulus change.

The present study starts from an established modal calibration and
does not yet define the full tolerance of the inspection chain to
transducer-coupling changes or drift between calibration and inspection.
In practice the nominal response could come from a reference specimen,
a previous inspection or a model-assisted local calibration, depending
on the transduction system. The fidelity required of that reference,
and the extent to which thickness and background material variation
can be estimated simultaneously from experimental data, remain to
be established. The sensitivity results above should therefore be
interpreted as the behaviour of the DEPTH measurement and inverse
once a usable modal calibration is available, rather than as a complete
robustness envelope for the calibration procedure itself. Experimental
validation with measured calibration repeatability and coupling variability
is therefore the next step.

\section{Conclusions\label{sec:conclusions}}

This paper has introduced DEPTH, a tomography formulation that uses
higher-order SH mode conversion to recover the axial and through-thickness
distribution of shear-modulus contrast. The central result of the
formulation is that each incident-received mode pair couples to at
most two cosine depth orders. Combining sufficiently diverse mode-pair
and frequency measurements therefore allows these known mixtures to
be separated in a compact depth basis, rather than treating through-thickness
variation as an additional unconstrained image dimension. In the weak-scattering
regime considered here, this depth encoding is supplied explicitly
by the known modal physics: the mode-pair kernel identifies the observable
through-thickness orders analytically and, after modal calibration,
the reconstruction remains a compact linear inverse rather than requiring
a general full-waveform optimisation. Although demonstrated here using
SH modes, the DEPTH concept is not intrinsically restricted to this
modal family: other guided-wave modes could also be used where their
mode shapes provide sufficiently diverse and well-characterised through-thickness
sensitivity functions.

The numerical evaluations support this principle under progressively
less idealised conditions. In the 10 mm two-dimensional plate, the
SH0-SH11 measurement set gives a through-thickness point-spread FWHM
of 0.84-1.03 mm and correctly localises four compact targets. The
single-surface 20-transmitter/20-receiver acquisition recovers the
same four targets from 4882 cross-mode measurements, showing that
direct access through the thickness is not required to obtain the
modal information. With full three-dimensional propagation and finite-width
transduction, all four finite ellipsoidal targets remain at the correct
axial and through-thickness locations and the reconstruction has a
whole-image correlation of 0.899 with the matched acquisition-limited
ideal result. A lower-surface patch, representing localised fatigue
damage, is also recovered at the correct surface. Measurement noise
produces gradual degradation at 30 dB, substantial degradation by
20 dB and poor localisation at 10 dB, while an isolated assumed-thickness
perturbation of magnitude up to 2\% after modal-dispersion calibration
leaves the recovered depth-varying coefficients largely unchanged
and retains a whole-image correlation above 0.971.

Two limitations are particularly important for practical use. The
surface and three-dimensional datasets considered here retain only
mode-converted measurements, so they recover the depth-varying contrast
relative to its local through-thickness mean; same-mode transmission
constrains only the global axial integral of that mean, so a local
mean field must be supplied by an additional axially localised measurement,
for example a scanned local pulse-echo travel-time measurement from
the same surface. Experimental implementation will also require modal
calibration that remains valid under transducer coupling, component
geometry and calibration-to-inspection drift, with wall thickness
either measured independently or constrained from the multimode dispersion.
The present forward operator also assumes a flat, nominally isotropic
reference waveguide; appreciable attenuation, curvature or background
anisotropy would require corresponding extensions to the calibrated
modal model.

Within these limits, the theory and numerical results establish a
basis for calibrated depth-resolved tomography of weak material-property
changes that would otherwise be diluted through the component thickness,
and, importantly, this has been demonstrated with practical, single-surface
access. DEPTH could ultimately support detection and quantification
of distributed material degradation, such as fatigue damage at a remote
surface, before it develops into a discrete crack or void, thereby
delivering critical information to industry.

\section*{Acknowledgements}

OpenAI ChatGPT (GPT-5.6 Sol) was used during manuscript preparation
to assist with literature searching, checking mathematical derivations
and scientific explanations, and revising text. The author directed
the use of the tool, independently verified the cited literature,
derivations, numerical results and final text, and takes full responsibility
for the work.

\section*{Data Availability}

The finite-element datasets supporting the numerical results in this
study are openly available at \linebreak{}
\url{https://doi.org/10.82186/9fpmx-hjs56}.

\appendix

\section{Derivation of the elastic mode-pair kernel\label{sec:appendix-kernel}}

The exact perturbed SH equation is
\begin{equation}
\nabla\cdot\left[\mu_{0}(1+q)\nabla(u^{\mathrm{i}}+u^{\mathrm{s}})\right]+\rho_{0}\omega^{2}(u^{\mathrm{i}}+u^{\mathrm{s}})=0.
\end{equation}
Since $L_{0}u^{\mathrm{i}}=0$, expansion gives
\begin{equation}
L_{0}u^{\mathrm{s}}+\nabla\cdot(\mu_{0}q\nabla u^{\mathrm{i}})+\nabla\cdot(\mu_{0}q\nabla u^{\mathrm{s}})=0.
\end{equation}

The last term is the perturbation acting on the scattered field and
is neglected under the first Born approximation, giving \eqref{eq:born}.
For incident mode $m$ take
\begin{equation}
u_{m}^{\mathrm{i}}=A_{m}\phi_{m}(z)e^{\mathrm{i}k_{m}x},
\end{equation}
and use the reciprocal received-mode field
\begin{equation}
v_{n}=B_{n}\phi_{n}(z)e^{-\mathrm{i}k_{n}x}.
\end{equation}

The received modal amplitude is proportional to the overlap of this
reciprocal field with the Born source,
\begin{equation}
\widetilde{d}_{nm}=-R_{n}\int_{\Omega}v_{n}\nabla\cdot(\mu_{0}q\nabla u_{m}^{\mathrm{i}})\,\mathrm{d}\Omega.
\end{equation}
Integration by parts gives
\begin{equation}
-\int_{\Omega}v_{n}\nabla\cdot(\mu_{0}q\nabla u_{m}^{\mathrm{i}})\,\mathrm{d}\Omega=\int_{\Omega}\mu_{0}q\nabla v_{n}\cdot\nabla u_{m}^{\mathrm{i}}\,\mathrm{d}\Omega.
\end{equation}

At the plate surfaces this boundary term vanishes because the reference
modes are traction free; at the axial limits the perturbation is assumed
to vanish outside the imaging region. The required derivatives are
\begin{equation}
\frac{\partial u_{m}^{\mathrm{i}}}{\partial x}=\mathrm{i}k_{m}A_{m}\phi_{m}e^{\mathrm{i}k_{m}x},\qquad\frac{\partial u_{m}^{\mathrm{i}}}{\partial z}=A_{m}\phi_{m}'e^{\mathrm{i}k_{m}x},
\end{equation}
\begin{equation}
\frac{\partial v_{n}}{\partial x}=-\mathrm{i}k_{n}B_{n}\phi_{n}e^{-\mathrm{i}k_{n}x},\qquad\frac{\partial v_{n}}{\partial z}=B_{n}\phi_{n}'e^{-\mathrm{i}k_{n}x}.
\end{equation}
Hence
\begin{equation}
\nabla v_{n}\cdot\nabla u_{m}^{\mathrm{i}}=A_{m}B_{n}\left[k_{m}k_{n}\phi_{m}\phi_{n}+\phi_{m}'\phi_{n}'\right]e^{\mathrm{i}(k_{m}-k_{n})x}.
\end{equation}
Collecting $R_{n}\mu_{0}A_{m}B_{n}$ and the experimental calibration
factors into $\widetilde{\kappa}_{nm}$ gives \eqref{eq:rawmeasurement}
and \eqref{eq:kernel}.

\section{Adjoint and coefficient-space regularisation\label{sec:appendix-adjoint}}

\subsection{Hermitian adjoint}

Starting from \eqref{eq:discreteforward} and the complex inner product,
\begin{equation}
\begin{aligned}\langle Aa,y\rangle & =\sum_{j}(Aa)_{j}^{*}y_{j}\\
 & =\sum_{p,l}a_{p,l}^{*}\left[\sum_{j}\Gamma_{jp}^{*}s_{j}y_{j}e^{-\mathrm{i}K_{x,j}x_{l}}\right].
\end{aligned}
\end{equation}
The expression in square brackets is therefore $(A^{H}y)_{p,l}$,
which gives \eqref{eq:adjoint}. This is the adjoint of the complex
forward operator; it is not an inverse.

\subsection{H1 regulariser}

For the expansion in \eqref{eq:qexpansion}, define
\begin{equation}
g_{p}=\int_{0}^{h}\cos^{2}\left(\frac{p\pi z}{h}\right)\,\mathrm{d}z=\begin{cases}
h, & p=0,\\
h/2, & p>0.
\end{cases}
\end{equation}
Using first differences in $x$, the axial contribution to the discrete
gradient penalty is
\begin{equation}
\sum_{p}\frac{g_{p}}{\Delta x}\sum_{l=0}^{L-2}(a_{p,l+1}-a_{p,l})^{2}.
\end{equation}
The corresponding rows of the regulariser may therefore be written
\begin{equation}
(L_{x}a)_{p,l}=\sqrt{\frac{g_{p}}{\Delta x}}\,(a_{p,l+1}-a_{p,l}).
\end{equation}
For the through-thickness derivative,
\begin{equation}
\frac{\partial q}{\partial z}=-\sum_{p>0}a_{p}(x)\frac{p\pi}{h}\sin\left(\frac{p\pi z}{h}\right).
\end{equation}
Orthogonality of the sine functions gives the depth contribution
\begin{equation}
\sum_{p>0}\Delta x\,\frac{h}{2}\left(\frac{p\pi}{h}\right)^{2}\sum_{l}a_{p,l}^{2},
\end{equation}
which can be represented by
\begin{equation}
(L_{z}a)_{p,l}=\sqrt{\Delta x\,\frac{h}{2}}\,\frac{p\pi}{h}a_{p,l},\qquad p>0.
\end{equation}
There is no depth term for $p=0$. The adjoint of $L_{x}$ is the
signed transpose of the first-difference incidence operator, while
$L_{z}$ is diagonal and self-adjoint. The implementation is checked
with its own real inner-product test.

\bibliographystyle{unsrt}
\bibliography{DEPTH-laptop}

\end{document}